\newcommand{\ignore}[1]{}

\documentclass[11pt]{article}
\usepackage[xdvi]{epsfig}
\newcommand{\half}{ {\scriptstyle \frac{1}{2} } }
\newcommand{\quarter}{ {\scriptstyle \frac{1}{4} } }
\newcommand\be{\begin{equation}}
\newcommand\ee{\end{equation}}
\newcommand\bea{\begin{eqnarray}}
\newcommand\eea{\end{eqnarray}}
\newcommand\nn{ \nonumber\\}

\newcommand{\text}[1]{\qquad \mbox{#1} \qquad}

\title{Hard Scattering in the M-Theory Dual for the  QCD String
\thanks{HET--1311: 
This work was supported in part by the Department of Energy under
Contracts No. DE-FG02-91ER40676 and No. DE-FG02-91ER40688}}

\author{Richard  C. Brower
\\ Physics Department\\Boston University\\
Boston, MA 02215, USA \\
\and
Chung-I Tan \\
Physics Department \\
Brown University\\
Providence, RI 02912, USA}

\begin{document}

\maketitle

\begin{abstract}
Conventional superstring amplitudes in flat space exhibits exponential
fall off at wide angle in  contrast to the  power law behavior found
in QCD. It has recently been argued by Polchinski and
Strassler~\cite{PolStras} that this conflict can be resolved via String/Gauge duality.
They carried out their analysis in terms of  strings in a deformed $AdS^5$
background. On the other hand, an equally valid approach to
the String/Gauge duality for 4-d QCD is based on M-theory in a
specific Black Hole deformation of $AdS^7
\times S^4$. We show that a very natural extension to this phenomenologically interesting
M-theory background also gives  the correct hard scattering power
laws.  In the Regge limit we extend the analysis to show the
co-existence of both the hard BFKL-like Pomeron and the soft Pomeron Regge
pole.
\end{abstract}

\newpage
\section{Introduction}

It is generally acknowledged that 't Hooft's $1/N_c$
expansion~\cite{tHooft} for QCD perturbation theory should map order
by order onto the topological expansion for some kind of ``QCD
string''.  Assuming confinement, the spectrum for the leading term at
infinite N must consist of an infinite sequence of stable glueballs
(e.g., closed string excitations) and the general form of unitarity
corrections at higher genus as well as non-perturbative effects, such
as instantons, skirmeons, have been studied extensively. However,
until recently with Maldacena's explicit examples of String/Gauge
duality~\cite{maldacena}, any direct relationship between a ``QCD
string'' and the fundamental superstring was missing. 

Most problematic for such a relationship is the stark contrast between
the exponentially soft properties of superstrings scattering at high
energies (in flat space) and the requirement from the leading large N
diagrams of QCD of hard partonic behavior at short distance. Thus the
super string appears to be in conflict with a large number of QCD
phenomena related to asymptotic freedom, scaling in deep inelastic
scattering, power law fall-off of form factors and wide angle
scattering to name a few. However in a most interesting paper, Polchinski
and Strassler~\cite{PolStras} may have begun to resolve this fundamental difficulty.
They have suggested a mechanism for hard wide angle string scattering
in the context of String/Gauge duality and this has been extended by
Polchinski and Susskind~\cite{PolSuss} to show how form factors for
strings may also exhibit power law behavior at large $Q^2$.

They argue that as strings move in the extra (radial) direction, they
exhibit both soft (IR) and hard (UV) aspects by appearing to the Yang
Mills observers as alternately ``fat" and ``thin" respectively. In
particular Polchinski and Strassler consider glueball scattering in
the dual description of  IIB strings scattering in an $AdS^5 \times
X_5$ background with a suitable IR cut-off (or deformation) in the
warped ``radial'' coordinate. They then relate the power counting
rules of Brodsky et al.~\cite{brodsky} for wide angle scattering and
form factors directly to the conformal scaling of the glueball wave
functions in the UV.  However there is another approach to 4d QCD,
based of M-theory (or IIA strings in an $AdS^7 \times S^4$ black
hole~\cite{wittenO}), which to date yields the most concrete results for QCD
glueballs masses~\cite{BMT}. Since the conformal scaling in $AdS^7$ is
different from the $AdS^5$ analysis, there is a question on how this
scenario can also agree with the parton counting rules for $QCD_4$.  

Here we show that if proper account is taken of the mapping that
relates the membrane in 11d M-theory to the 10d IIA string theory, new
scaling factors for the effective string parameters result which
correct conformal scaling powers to again reproduce the parton result.
While this result is certainly expected on the basis of String/Gauge
duality, it is useful to check it and to understand how it comes
about.  Section 2 explains how the parton scaling for glueball
operators in $AdS^5$ (IIB string) and in $AdS^7$ (M-theory) are
reconciled. In Section 3 the full form for the wide angle scattering
amplitude in M-theory is derived and the cross sections are compared
with the weak coupling parton~\cite{brodsky} and the strong coupling $AdS^5$
~\cite{PolStras} results. In Section 4 we consider Regge behavior in
the near-forward limit in order to clarify the relation between hard
BFKL Pomeron~\cite{BFKL} and the soft Pomeron responsible for the
Reggeized glueball exchange process at large N. We end in Section 5
with a brief comment on issues related to the high energy Froissart
bound.

%\newpage
\section{Wide Angle Glueball Scattering in the M-theory}

In QCD wide angle scattering exhibits a power law fall-off in energy up to
small logarithmic corrections due to asymptotic freedom. For example the 2-2
glueball amplitude scales as \be A_{qcd}(s,t) \sim
\Big(\sqrt{\alpha'_{qcd}} p \Big )^{4-n} \ee at large $p = \sqrt{s}$ and fixed
$-t/s$.  The power is determined by $n= \sum_i n_i$, where $n_i$ is the
number of ``partons''  for the external bound state
or more precisely lowest twist $\tau_i = d_i - s_i$
for the interpolating fields.  Here we focus on
this power behavior for each external line to explain how M-theory (or
$AdS^7$/gauge duality) manages in the end to give the same power as that
found earlier by Polchinski and Strassler~\cite{PolStras} for $AdS^5$
String/Gauge duality for $QCD_4$, postponing to the next section the full
derivation of the scattering cross section.

The essential observations in Ref~\cite{PolStras} are as follows. Glueball
scattering amplitudes in the gravity description are given by the product of
wave functions, $\phi_i(r,X) \exp[ix_ip_i]$ for each external leg and a local
string scattering amplitude $A(p_i,r,X)$ in the bulk gravity theory. They
take the bulk metric to be $AdS^5 \times X_5$,
\be
 ds^2 = \frac{r^2}{R^2} \eta_{\mu\nu}dx^\mu dx^\nu +
\frac{R^2}{r^2} dr^2 + R^2 ds^2_X \; , 
\ee
with $ds^2_X$ denoting an additional 5 compact dimensions.  R is the ads
radius.  By introducing an IR cut-off ($r > r_{min}$) the string states (aka
glueballs) are discrete and massive.  In a plane wave state ($\exp[i x p]$)
at fixed external momentum $p$, the proper distance ( $\Delta s \simeq (r/R )
\Delta x$) is red shifted in the IR (small r). Alternatively one may view the
local scattering to take place with an anti-red-shifted effective momentum,
\be
\hat p_s(r) = \frac{R}{r} p \; .
\ee
At high string momentum ($l_s p_s > 1$) relative to the string scale, $l_s =
\sqrt{\alpha'_s}$, wide angle scattering is indeed exponentially suppressed
because of the Hagedorn-like spectrum of soft modes at high
energy. Consequently the wide angle scattering of the external
glueballs is negligible except in a small region in the IR,
\be
r > r_{scat} = \sqrt{\alpha'_s} R p \; ,
\ee
that shrinks as a function of momentum. The dominant contribution to the wide
angle scattering therefore scales due to UV boundary condition on the wave
function. This scaling rule is set by the conformal weight of the
corresponding gauge operator dual to the string state.
\be
\phi_i(r_{scat})  \sim 
\Big (r_{scat}/r_{min}\Big )^{-\Delta^{(i)}_4} 
\sim
\Big (\sqrt{\alpha'_s} R p /r_{min}\Big )^{-\Delta^{(i)}_4} 
\sim    ( \sqrt{\alpha'_{qcd}}\; p )^{- \Delta^{(i)}_4}
\label{eq:ads5}
\ee
For $AdS^5$, the conformal dimension $\Delta^{(i)}_4$ is equivalent to the
twist $n^{(i)}$ required by correspondence to the power law fall-off for hard
parton scattering.  Thus it is a consequence of String/Gauge duality that the
color singlet string description encodes the gauge theory parton constituent
rule.  The final step in Eq.~\ref{eq:ads5}, converting to the hadronic string
scale, comes from the relation 
\be
 \alpha'_{qcd} \sim (R/r_{min})^2 \alpha'_s\;.
\label{eq:alpha5}
\ee
The 2nd power in $R/r_{min}$ reflects the stabilization of minimal area
at the IR cut-off for a world sheet spanning the Wilson loop.

How does this work in M-theory?  One begins in M-theory  with an 11d Black Hole
deformation of $AdS^7 \times S^4$,
\be
ds^2 = \frac{r^2}{R^2} \eta_{\mu\nu}dx^\mu dx^\nu + \frac{R^2}{r^2}(1-
r_{min}^6/r^6)^{-1} dr^2 + R^2 ds^2_Y \; ,
\ee
which is essentially the same form as $AdS^5 \times X_5$ except for an
explicit IR cut-off and an extra 6, instead of 5, compact
coordinates. Again we treat wide angle scattering as a local event in
the bulk with the same red-shift factor rescaling to the local
scattering momenta, $\hat p_s(r) =( R/r) p $, as before.  However this
alone would lead to the wrong parton scaling for hard scattering cross
sections, because the $AdS^7$ asymptotic form for glueball wave
functions, $\phi_i(r) \simeq C_i (r/r_{min})^{-\Delta^{(i)}_6}$, are
not the same as conformal power for $AdS^5$ (e.g., a scalar glueball
has $\Delta_6 = 6$ instead of $\Delta_4 = 4$).  To avoid this
consequence, one must realize that from an M-theory perspective,
strings are a consequence of membranes wrapping the ``11th'' dimension
and that in $AdS^7$ this 11th dimension is warped just like another
spatial coordinate ($x^\mu$) with the proper size radius: $\hat
R_{11}(r) = (r/R) R_{11}$.  To account for this effect one must
introduce local values for the effective string length and coupling
constant:
\be
\hat l^2_s(r) = \frac{R}{r} (l^3_p / R_{11})\;,  \quad {and} \quad  
\hat g^2_s(r) = \frac{r^3}{R^3} (R^3_{11}/l^3_p)\;.
\ee
(Our convention is to scale the local variable relative to the value
at the AdS radius, so that $ \hat g_s(R) = g_s$ and $\hat l_s(R) = l_s
= \sqrt{\alpha'_s}$ at $r = R \; .$)  This additional deformation is
precisely what is required\footnote{An alternative approach is to go
directly to 10d IIA string theory in the string frame. Here one sees a
rescaling by $r^{3/2}$ for the local momenta, which is equivalent to
the rescaling of $l_s(r) p_s(r)$ in our 11d approach. We have found this
correspondence less intuitive}.

The new definition of the scattering region at wide angles,
\be
r > r_{scat} = \hat l_s(r_{scat}) R\; p = \sqrt{\alpha'_s} \; R^\frac{2}{3} \;
r_{scat}^{-\half} \; p\; ,
\ee
leads to 
\be
\phi_i(r) \sim \Big (r_{scat}/r_{min} \Big )^{-\Delta^{(i)}_6} 
\sim  \Big ( \sqrt{\alpha'_s}\; p/ \sqrt{ r^3_{min}/R^3} \Big)^{-\frac{2}{3} \Delta^{(i)}_6}
\sim  \Big( \sqrt{\alpha'_{qcd}}\; \Big)^{-\frac{2}{3} \Delta^{(i)}_6}
\label{eq:ads7}
\ee
for each external line.  For example for the $0^{++}$
scalar glueball corresponding to interpolating YM
operator $Tr[F^2]$,  the factor of $2/3$ exactly compensates for the the
shift in the conformal dimension from $\Delta_4 = 4$ for $AdS^5$ to
$\Delta_6 = 6$ for $AdS^7$ as it should to give the parton results,
$n_i = 2\Delta^{(i)}_6/3$.  This time, in converting to the hadronic scale in
Eq.~\ref{eq:ads7}, we must realize the relationship of $\alpha'_{qcd}$
to the string scale is
\be
\alpha'_{qcd} \sim (R/r_{min})^3 \alpha'_s \; ,
\ee
which differs from the $AdS^5$ string relation (\ref{eq:alpha5}).  The
3rd power is a consequence of the fact that in M-theory the area law
for the Wilson loop really comes from a minimal volume for a wrapped
membrane world volume stabilized at $r \simeq r_{min}$ rather than a
minimal world surface area for a string. In summary the requisite
adjustment is essentially a multiplicative factor of $2/3$ reflecting
the difference between scattering strings with 2d world sheets by
membranes with 3d world volumes.

%\newpage
\section{Scattering Amplitude}

We now look at the full scattering amplitude in more detail for 
the M-theory construction. Here we need the 
precise form $AdS^7 \times S^4$ Black Hole metric,
\be
ds^2 = \frac{r^2}{R^2} (\eta_{\mu\nu}dx^\mu dx^\nu + dx_{11}^2) +
\frac{R^2}{r^2}(1- r_{min}^6/r^6)^{-1} dr^2 +
\frac{r^2}{R^2}(1- r_{min}^6/r^6) d\tau^2   +\quarter R^2 d^2 \Omega^2_4\; ,
\ee
where $\half R d\Omega_4$ are  coordinates for $S^4$ with 1/2 the
AdS radius (R), $dx_{11}$ on an $S^1$ circle with radius $R_{11}$ and
$d\tau$ on a ``thermal'' $S^1$ circle ($\tau \leftrightarrow \tau +
\beta$).  After these compactification, the boundary (2,0) CFT for
$AdS^7$ reduces to 5d Yang-Mills theory at finite temperature,
$\beta^{-1}$.  Assuming anti-periodic boundary condition in $\tau$,
all conformal and SUSY symmetries are broken with $\Lambda \equiv
\beta^{-1}$  setting the scale for glueball masses in
(strong coupling) 4-d QCD. (The requirement that the horizon is a
coordinate singularity fixes $\beta = (2 \pi/3) (R^2/r_{min})$.)

The $2$-to-$m$ glueball scattering amplitude $T(p_i)$ in a
gravity-dual description, is expressed in terms of the bulk scattering
amplitudes  $A( p_i, r_i,Y_i)$ for the plane wave glueball
states $\phi_j(r,Y) \exp[i x_j p_j]$  associated
with each external line:   
\be
T(p_i)= \prod_j \int d\mu_j \; \phi_j(r_j,Y_j)
\; \; A(p_1, r_1, Y_1,\cdots, p_{m+2},r_{m+2},Y_{m+2}) \; .
\ee
The transverse volume element in the bulk is
$$\int d\mu = \int_{r_{min}}^{\infty} dr
\int_0^{R_{11}}dx_{11}\int_0^{\beta} d\tau \int d\Omega {\sqrt {-g}},
$$
with $\sqrt{-g} = {r}^5/(16 R)$.  At wide angle we approximate the scattering
by a local ``form factor'' $A(p_i)$ which is exponentially suppressed for
momentum large with respect to the local string scale. Thus we can
approximate the scattering amplitude by \be T(p_i)=\frac{R_{11}\beta }{R}
\int_{r_{min}}^{\infty} dr r^5 A(p_i) \prod_{i=1}^{m+2} \phi_i(r) ,
\label{eq:local}
\ee
in the region where there is appreciable wide angle scattering,
\be
l_s(r) \hat p_s(r) \leq  O(1) \quad {or} \quad r^3 \geq r^3_{scat} \equiv 
  R^3 \alpha'_s  p^2  \;.
\ee
Obviously one would like to be able to justify this approximation with more
detailed calculations.

The membrane scattering amplitude at large r is a constant which by
dimensional analysis can only be a power of $l_p$, 
\be 
A = Z_0 \;\;
l_p^{\frac{(D-2) m}{2}} \frac{1}{l_p^2}\;.
 \ee 
Each of the m additional external
legs carries a factor $l_p^{\frac{9 }{2}} = \sqrt{R_{11}} g_s {\alpha'}_s^2 $,
consistent with the scattering of 10d strings. In flat space we know that
the reduction from 11d M-theory to 10d strings requires an additional
factor $Z_0 = l^2_p/l^2_s$.  In $AdS^7$ the local conformal factor is
$$l^2_p/l^2_s(r) \sim \exp[{\frac{2 \phi_0(r)}{3}}] \; ,$$
 which takes you to the string frame.  For glueball scattering this
factor is characterized by its value at $r_{GB} \sim r_{min}$ where
the glueball wave functions have there largest support,
\be
Z_0 = l^2_p/l^2_s(r_{min}) = (g^2_{YM} N)  / N^{2/3} \; .
\ee
The standard normalization condition in 11d fixes the parametric
dependence of the normalization constant in the asymptotic behavior of
the glueball wave functions, $\phi_i(r) \simeq C
(r_{scat}/r_{min})^{-\Delta^{(i)}_6}$, to be $C^{-2} = R_{11}\beta R
r_{min}^4 = R_{11} (R^3\Lambda)^3$.

Thus assembling all factors, the contribution from the large r region is
\be
T(p)= (Z_0R_{11}\beta/R l_p^{11}) (l_p^{9} C^2)^{(m+2)/2}  \int_{r_{scat}}^{\infty} dr r^5\Big ( \frac
{r_{min}}{r}\Big ) ^{\Delta_6} \; ,
\ee
where $\Delta_6 \equiv \sum_{i=1}^{m+2} \Delta^{(i)}_6$, and the $r$-integration
region has also been restricted to $r_{scat}< r$. After performing the
integral the result is 
\be
T(p_i) = \frac{\Big(\sqrt{\alpha'_{qcd}}\Big)^{m-2}}{N^m } \Big (
\sqrt{\alpha'_{qcd}}\> p\Big ) ^{-(\frac{2}{3} \Delta_6 -4)} \; ,
\ee
expressed entirely in terms of the hadronic scale $\alpha'_{qcd}$. As
we mentioned above for scalar glueballs $\frac{2}{3} \Delta_6 = 4$,
the number of constituent partons in the lowest dimension
interpolating fields (e.g.  $Tr[F^2]$), in general this is the
``twist'' $\tau_i$. We have checked explicitly that this works for all
the tensor and vector glueball states identified in strong coupling in
Brower, Mathur, and Tan \cite{BMT}.  One must make use of the general
formula for conformal dimensions, $\Delta_d^{(p)} = d/2 + \sqrt{(d/2)^2 +
m^2} -p$, for tensor fields in $AdS^{d+1}$ to check that the rescaling
of the exponent by $2/3$ from $AdS^5$ to $AdS^7$ is a general feature.

Note that the first two factors correctly reproduce the leading N
behavior, $N^{-m}$, and the proper dimensional dependence,
$\Lambda^{-m+2}$.  Note we never needed the explicit expression for
the $\alpha'_{qcd}$.  In the strong coupling limit the actual value is
\be
\alpha'_{qcd}\equiv   {\frac {27}{32\pi^2}} (g^2_{YM} N  \Lambda  ^{2})^{-1},
\ee
where we have use the definition of the Yang Mills coupling $g^2_{YM}
= g_s l_s \Lambda$ and the expression for the $R^3 = N l^3_p$ at
large N.

\subsection{ Comparison with $AdS^5$ and QCD parton cross sections}

It is interesting to compare the resulting M-theory fixed-angle differential
cross sections with that from $AdS^5$ as well as that from lowest order
perturbative QCD.  Consider a $2$-to-$m$ process, $p_1\> + \>p_2
\rightarrow p_3\> +\cdots\cdots+p_{m+2}$, in the high energy fixed-angle
limit. In the CM frame, with $s\equiv (p_1+p_2)^2\rightarrow \infty$ , all
components of particle momenta are of the order $\sqrt s$ and all ratios
$p_i\cdot p_j/s$ are fixed. For each particle in the final state, let
$x_i= E_i/\sqrt s$ and denote the direction of its three-momentum by
$\Omega_i$. Due to overall energy-momentum conservation, we only need to
consider independent variables $\{x_i, \Omega_i\}$, $i=3,\cdots, m+1$.  The
resulting differential cross section can be written as
$\Delta\sigma_{2\rightarrow m}\equiv {d\sigma}/{d x_i d \Omega_i} \simeq
s^{m-3} |T|^2  \; .$

For QCD hard scattering in lowest order perturbation theory, this
cross section is given by
\be
\Delta \sigma_{2 \rightarrow m} \simeq \frac{1}{s}\;f(\frac{p_i \cdot p_j}{s}) 
 \frac{(g^2_{YM} N)^m}{N^{2m}} \;\prod_i \Big( \frac{g^2_{YM} N \Lambda^2_{qcd}}{s} \Big)^{n_i - 1}\;,
\ee
where $f$ can in principle be calculated. This serves as the ``Born"
term which will be modified when asymptotic freedom is taken into
account.  The corresponding $AdS^5$ cross section for $AdS^5$ at strong coupling is 
\be
\Delta \sigma_{2 \rightarrow m} \simeq \frac{1}{s}\;f(\frac{p_i \cdot p_j}{s}) 
 \frac{(\sqrt{g^2_{YM} N})^m}{N^{2m}} \;\prod_i \Big( \frac{\sqrt{g^2_{YM} N}
 \Lambda^2_{qcd}}{s} \Big)^{n_i - 1} \; .
\ee
while for M-theory our result can be expressed 
as
\be
\Delta \sigma_{2 \rightarrow m} \simeq \frac{1}{s}\;f(\frac{p_i \cdot p_j}{s}) 
 \frac{1}{N^{2m}} \;\prod_i \Big( \frac{1}{\alpha'_{qcd} s} \Big)^{n_i
 - 1} .
\ee
The latter two  strong coupling results have the same power law term if we make
use of the appropriate expression for the QCD string tension at
strong coupling,
\bea
1/\alpha'_{qcd} &\sim& (g^2_{YM} N)^\half \Lambda^2_{qcd}  \quad \mbox{for $AdS^5$ IIB strings}\;, \nn 
1/\alpha'_{qcd} &\sim& g^2_{YM}\; N  \Lambda^2_{qcd}  \qquad \mbox{for $AdS^7$ M-theory}\; . \nonumber
\eea
Of course the explicit dependence on $g^2_{YM} N$ for $AdS^5$ and
$AdS^7$ need not agree with each other or with weak coupling QCD due
to non-universal artifacts at strong coupling.  Each of them can be
``converted'' into the perturbative result by the following
substitution rules, 
\bea g^2_{YM} N \leftrightarrow & (g^2_{YM} N)^\half \; &\mbox{for} \quad AdS^5
\;,\nn (g^2_{YM}N)^m \leftrightarrow& 1 \; &\mbox{for} \quad AdS^7 \; ,\nonumber
\eea 
after $\Delta \sigma$ is expressed in terms of the appropriate expression
for $\alpha'_{qcd}$. It was noted in Ref~\cite{PolStras} that the
substitution rule for $AdS^5$ works in several instances comparing
strong and weak coupling results. We find it interesting that the 
$AdS^7$ cross section is naturally expressed entirely in term of the
fundamental string tension for QCD. With dimensional transmutation this
is also true of QCD itself at large N.  Perhaps this is also more general.

These results can also be interpreted probabilistically. The cross
section is given by the product of a probability ${\cal P}_{m+2}$ of
finding all partons within each hadron in a transverse area of order
$\Delta A_\perp \sim 1/s$ and an elementary ``Rutherford" cross
section $\Delta \sigma_0$,
$$\Delta \sigma_{2 \rightarrow m}= {\cal P}_{m+2}\; \Delta \sigma_0 \;.$$ 
In all three cases, the probability ${\cal P}_{m+2}$ is given
geometrically as $\prod_i \Big(1/{\alpha'_{qcd} s} \Big)^{(n_i - 1)}$.
The Rutherford cross section, in additional to a factor of
${N^{-2m}}$, is again geometrical, $\Delta \sigma_0 \sim (C/s)
f(\frac{p_i \cdot p_j}{s}) {N^{-2m}} $.  Only the non-universal
prefactor distinguishes the three cases: For QCD, $AdS^5$ and $AdS^7$
the prefactors are  $C= (g^2_{YM}N)^m,\; (g^2_{YM}N)^{m/2}$ and $1$ 
respectively.

%\newpage
\section{Near-Forward Scattering and Regge Behavior}

High energy hadron phenomenology has revealed that, in the
near-forward limit, the elastic scattering amplitude contains both ``hard"
and ``soft" components~\cite{LT}. Based on a perturbative QCD analysis,
valid in the limit $s>>|t|>> {\alpha'_{qcd}}^{-1}$, contribution to
high energy hadronic total cross section from ``hard collisions" can
be evaluated, leading to a ``hard Pomeron" exchange, the celebrated
BFKL Pomeron~\cite{BFKL}. On the other hand, phenomenological
analysis, supported by large-N QCD string picture, demonstrates that
scattering at large impact parameter is dominated by the exchange of a
``soft'' Pomeron Regge pole.  The key distinguishing feature is their
Regge slopes.  The slope for the hard Pomeron is small; it is in fact
a fixed branch point in the complex angular momentum plane if the
running in QCD coupling is not taken into account. In contrast, the
soft Pomeron is a factorizable Regge pole, having a normal hadronic
slope with its first Regge recurrence at the $2^{++}$ tensor
glueball. (Experimentally, one has $\alpha'_{P}\sim .2 \sim .3 \> $
$GeV^{-2}$ for the soft Pomeron.)  We shall demonstrate next that both
these components emerge naturally in our gravity-dual description for
the leading large-N contribution to QCD.

To be specific, consider the 2-to-2 scattering, $p_1\> + \>p_2
\rightarrow p_3\> +\> p_4$, with Mandelstam variables $s\equiv
(p_1+p_2)^2$ and $t\equiv (p_3-p_1)^2$ in the Regge limit
$s\rightarrow \infty$ at $t$ fixed.  The assumption of a single local
scattering, Eq.~\ref{eq:local}, leads to $ T(s,t) =
\int_{r_{min}}^{\infty} dr \;{\cal K}(r) \> A(s, t, r), $ where $A$ is
a local four-point amplitude, and ${\cal K}(r) \sim r^5 \phi_1(r)
\phi_2(r)
\phi_3(r) \phi_4(r)$, up to a constant. Converting to local string
parameters, amplitude $A(s, t, r)$ depends only on $\alpha'_s \hat s(r)$ and
$\alpha'_s \hat t (r)$, which in the Regge limit becomes
\be 
T(s,t) = \int_{r_{min}}^{\infty} dr\; {\cal K}(r) \> \beta( \hat t)
(\alpha'_s \hat s)^{2 + \alpha_s' \hat t}\;,
\label{eq:regge}
\ee
with t-channel trajectory,
\be 
\alpha_s(\hat t ) = 2 + {\alpha'_s}\> \hat t\;. 
\ee 
The local Mandelstam invariants are defined by $\hat s (r)= (R/r)^3 \>
s \>$ and $\hat t (r) =(R/r)^3 \> t \>.$ The 3rd power can be
understood from our earlier discussion by noting that $\alpha'_s \hat
s(r) = \hat l^2_s(r) (\hat p_1(r) + \hat p_2(r))^2$.  For $AdS^5$ only
the momentum is rescaled, leading to 2 power of $(R/r)$.  Note that
the Regge slope is $\alpha_s'$ and its intercept, $\alpha_s(0)=2$,
would normally correspond at $t=0$ to a spin-2 graviton exchange.
However, due to the confinement mechanism giving rise to glueball
masses, this intercept is shifted lower so that the trajectory
intercept in less than 2.  (More on this point shortly.)

Since $\hat t$ is in the interval $[ (R/r_{min})^3 t, 0]$, one finds
that $T(s,t)$ in the Regge limit is given by a linear superposition of
power-behavior in $s$ averaged over the radial coordinate, i.e., a cut
in the J-plane.  In the near-forward limit where $|t|$ is small the
cut is localized near $j = 2$ with the high energy behavior
approximated by 
\be T(s,t)\sim f(t) s^{<\alpha_s(\hat t)>}, \ee
where $<\alpha_s(\hat t)>\simeq \alpha_s(0)=2$. The precise result for the
leading high energy behavior can be found
by a saddle-point analysis or equivalently by studying the
Mellin transform to the J-plane,
\be 
T(j,t) = \int_{r_{min}}^{\infty} dr{\cal K} (r) \Big( \frac
{r_{min}}{r}\Big)^{-(j+1)} \frac {\beta( \hat t(r))}{j-2+ \alpha'_s(R/r)^3)t
  }.  
\ee
For $t\neq 0$, there is a pair of branch points due to end-point
pinching.  From the upper limit, one finds a branch point located at
$j=2$. The large r (UV) behavior for the product of wave functions
leads to $T(j,t)\sim f(t) (j-2)^{\gamma}$, where $\gamma=
\Delta/3-2>0$.  From the lower limit (IR), since wave functions are
$0(1)$, one finds a logarithmic branch point at $j=2+\alpha'_s
(R/r_{min})^3 t $.

For $t<0$, the leading asymptotic behavior at infinite $s$ comes from
the UV region near the $AdS$ boundary, $T(s,t)\sim f(t)
s^{\alpha_s(0)}/{(\log s)}^{\gamma+1}$, as in the case of the
fixed-angle scattering.  So it should also be identified with hard
gluon effects and therefore represents the physics of the BFKL hard
Pomeron.  Moreover, due to the local approximation in the bulk AdS
string scattering, this does not lead to a factorizable t-channel
exchange, another feature of the BFKL Pomeron. To properly isolate the
hard processes one should introduce a cutoff, $r_h$, where
$r_h>>r_{min}$.  The Born term of the $BFKL$ hard Pomeron in a
gravity-dual description therefore corresponds to the branch cut in
the interval close to j=2, $[2+\alpha'_s(R/r_h)^3 t,2]$.  Exactly at
$t=0$, these two branch points coincide and the hard process is give
by pole, $T(j,0) \simeq 1/{(j-2)}$.  For small $t\simeq 0$, the branch
cut acts effectively as a single pole, with a small effective slope:
\be 
\alpha'_{BFKL}(0)\sim
(R/r_h)^3\alpha'_s= (r_{min}/r_h)^3\alpha'_{qcd}<< \alpha'_{qcd}.  
\ee
Exchanging a BFKL-Pomeron naturally leads to a diffraction peak for the
elastic differential cross section.  Going to the coordinate space, one finds,
for a hard process, the transverse size is given by 
\be 
<{\vec X}^2> \sim (r_{min}/r_h)^3 \alpha'_{qcd} \log s +{\rm constant} \; . 
\label{eq:xsq} 
\ee 
If the cutoff, $r_h$, which characterizes a hard process, increases
with $s$, e.g. $r_h^3\sim \log s$ or faster, transverse spread would
stop.  In the language of a recent study by Polchinski and
Susskind,~\cite{PolSuss} this corresponds to ``thin" string
fluctuation. 

Having identified the hard Pomeron, the question remains: where is the soft
Pomeron? At moderate values of s we may imagine that the dominate scattering
occurs in the IR region where the glueball wave-function are large.  Thus in
Eq.~\ref{eq:regge}, a large contribution comes for the integrand with $r_{GB}
\simeq r_{min}$,
\be 
T(s,t)\sim A( s, t, r_{GB}) 
\sim (\alpha'_{qcd} s )^{2+\alpha'_{qcd}t}/\log s.  
\ee
where we have made use of the fact that $\alpha'_s \hat s (r_{min},s) =
\alpha'_{qcd} s\>$. The effective Regge slope is typical of a soft Pomeron. Of
course this is a rough characterization.  Actually the integral does
not yield a pure pole (i.e. power behavior), but more important it
does not even factorize in the t-channel because of our local
scattering approximation. In reality we know that at large N, there is
a infinite spectrum of stable glueballs that propagate on shell and on
average gives the cut in the s-plane, $(-s)^{\alpha_{P}(t)}$. In the
old language of dual models, this is dual to the t-channel Regge
pole. Again by continuing to integer J these are on shell modes. All
of this contradicts a local scattering picture for point like object
in bulk AdS. One needs much more powerful methods to find the on-shell
string spectrum to really understand these issues
completely. Nonetheless consistent with the known spectrum of
glueballs at strong coupling we can anticipate that the IR-region must
give a factorizable Regge pole contribution,
\be 
T(s,t)\sim A( s, t, r_{min}) \sim (\alpha'_{qcd} s ) ^ {\alpha_{P}(t)},
\ee
where 
\be 
\alpha_{P}(t) = \alpha_{P}(0) + \alpha'_{qcd} t\>.  
\ee
Note that the slope of this Regge trajectory is $\alpha'_{qcd}$ and the first
physical particle lying on this trajectory is a tensor glueball, i.e.,
$\alpha_{P}(m_T^2)=2$.  For elastic scattering, this Regge trajectory should
be identified with the ``soft Pomeron".  That is a Pomeron at $t=0$ which is a 
pole
with slope renormalized from the string scale to a hadronic scale: $\alpha'_s
\rightarrow\alpha'_{qcd}$.  From the diffraction peak for the elastic
differential cross section, one finds that the transverse size increases as
\be <{\vec X}^2> \sim \alpha_{qcd}' \log s+{\rm constant}.  \ee 
It is interesting to note that this ``divergence" in hadron size, in
the infinite $s$ limit, corresponds precisely to the divergent
``transverse size" in a string picture recently considered by
Polchinski and Susskind.  For the form factor this divergence must be
removed by fluctuations into the thin string (UV) regime to avoid the
traditional disaster of an infinite rms radius for hadronic form
factors in the string description.

A rigorous treatment of the Regge behavior in the leading large-N
approximation should give a modified four-point function (or Veneziano
amplitude) with both soft and hard Pomerons combined. The 
J-plane of the improved Veneziano amplitude must have contributions
from scattering through configurations involving both thin and fat
strings~\cite{PolSuss}. This is consistent with a partonic picture
for which it has been suggested in the past that these contributions can serve 
as
``Born terms" of an iterative sum, leading to a single ``Pomeron"
which captures the physics of both hard scattering at short-distance
as well as infrared physics of confinement at large
distance~\cite{LT}. On the other hand, phenomenologically it appears
that for current accelerator energies it is sufficient to treat these
contributions additively. Nevertheless, it is a major challenge to be
able to really understand these effects entirely within the color
singlet formulations of perturbative string field theory.

%\newpage
\section{Discussion}

Let us conclude with some comments on the nature of Pomeron as well as
some speculative remarks on the constraint of unitary beyond the
topological string expansion. Since the $1/N$ expansion is a
perturbative solution to unitarity, at any finite order it needs not
obey the non-linear unitarity bounds.  It has been noted that the
naive power law fall off for the wide angle 2-to-m cross
sections~\cite{brodsky} of the parton model saturates, but does not
violate, the unitarity bound.  However in the Regge limit, our
calculation gives both hard and soft Pomeron contributions with a
common intercept greater than one,
$$\sigma_{total}\sim C \; s^{\Delta}\;,$$
for $\alpha_P(0) = 1+ \Delta$ in contradition to the Froissart bound,
$\sigma_{total} \le C (\log s)^2$ asymptotically.  Indeed this picture
is supported experimentally by power law with an intercept,
$\alpha_P(0) = 1+ \Delta$, where $\Delta \simeq 0.08$, while at
presently available energies (e.g., $p\bar p$ scattering up to TeV
range) the cross sections are still an order of magnitude smaller than
the absolute unitarity bound.

From the large N perspective this power appears to be controlled by
low energy dynamics, i.e., by the requirement that the trajectory
interpolates the lightest tensor glueball, $\alpha_P(m_T^2)=2$.  We
have previously pointed out~\cite{BMT},  if  one
assumes a linear trajectory and the strong coupling approximation
to the tensor glueball mass, one obtains
$$\Delta=\alpha_P(0)-1\simeq 1-0.66(\frac{4\pi} {g^2_{YM} N}) \; ,$$
in contrast to $\Delta= 1- O(1/ \sqrt{g^2_{YM} N })$ in a
deformed $AdS^5$ background~\cite{PolStras,PolStrasDIS}. In both cases, Pomeron can be interpreted
as ``confined graviton", with $\alpha_P(0) < 2$.  If one also accepts
fits to the empirical value of $\Delta$, this relation is consistent
with the lattice estimate of the bare coupling constant at the weak to
strong crossover. Thus a crude matching between weak and strong
coupling seems to make sense.~\cite{BMT}

More importantly, with $\Delta>0$, Pomeron-like power behavior leads
to violation of elastic unitarity for partial wave amplitudes.
Traditionally one imagines restoring unitarity at high energies via an
s-channel iterations such as the eikonal mechanism which requires
summing higher genus contributions to all orders. In such a scheme, a
``disk-like" picture emerges, with an effective hadronic radius
$$
R_{eff}\sim \sqrt {\alpha'_{qcd} \Delta} \log s \; ,
$$
obeying the Froissart bound.

More recently, Giddings~\cite{Giddings} has also addressed the issue
of unitarity violations at much higher energies where
non-perturbative string interactions must be taken into account,
e.g., the effect of Black Hole production.  Curiously the production
cross section for a single Black Hole leads to a cross section
increasing with $s$ as a power, $\Delta=1/(D-3)$.  With $D> 4$, this
is consistent with our bound, $\alpha_P(0)\leq 2$.  Giddings goes on
to ``unitarize'' this perturbative black hole result to saturate the
Froissart bound. It is then interesting to raise the question of
whether the Pomeron intercept is constrained by a matching condition
to the Black Hole production and what is the Gauge correspondence for
this unitarization mechanism.

Much more can be said concerning scattering in the near-forward
limit~\cite{LT,Others} but we shall defer that to a future
publication.  In summary, in spite of these insights on high energy
processes for a QCD string, a deeper microscopic understanding of
these issues in terms of String/Gauge duality is lacking.  In
particular as emphasized in a recent paper~\cite{PolStrasDIS}, scaling
in the deep inelastic limit requires a much more explicit connetion to
individual partons. Here these simple QCD-like models lead to
structure funtions but the scaling laws for the cross sections fail by
powers relative to QCD.  The recent analysis of the Penrose limit of
$AdS^5 \times S^5$ type IIB string theory dual to ${\cal N} = 4$ SUSY
Yang-Mills by Berenstein, Maldacena and Nastase~\cite{bmn} and high J
configurations by Gubser, Klebanov and Polyakov~\cite{gkp} are
promising developments that one may hope will lead to a rigorous
derivation the String/Gauge duality for the partonic properties in
this context.

{\bf Acknowledgement}: We would like to thank A. Jevicki, D. Lowe, and M. 
Schvellinger for discussions and comments.

\end{document}